\documentclass[12pt,american,canadian,french,english,authoryear, 12, preprint]{article}
\usepackage[T1]{fontenc}
\usepackage[latin9]{inputenc}
\usepackage{geometry}
\usepackage{babel}
\makeatletter
\addto\extrasfrench{%
   \providecommand{\fg}{\ifdim\lastskip>\z@\unskip\fi~\frqq}%
}

\makeatother
\usepackage{array}
\usepackage{textcomp}
\usepackage{multirow}
\usepackage{amsmath}
\usepackage{graphicx}
\usepackage{setspace}
\usepackage[authoryear]{natbib}
\usepackage[unicode=true]
 {hyperref}

\makeatletter

\providecommand{\tabularnewline}{\\}

\newcommand{\lyxaddress}[1]{
\par {\raggedright #1
\vspace{1.4em}
\noindent\par}
}

\usepackage[left,modulo]{lineno}

\@ifundefined{showcaptionsetup}{}{%
 \PassOptionsToPackage{caption=false}{subfig}}
\usepackage{subfig}
\makeatother

\begin{document}

\title{Multifractal topography of several planetary bodies in the Solar
System}

\author{François Landais (1), Frédéric Schmidt (1), Shaun Lovejoy (2)}
\maketitle

\lyxaddress{(1) GEOPS, Univ. Paris-Sud, CNRS, Université Paris-Saclay, Rue du
Belvédère, Bât. 504-509, 91405 Orsay, France (2) McGill University}
\begin{abstract}
Topography is the expression of both internal and external processes
of a planetary body. Thus hypsometry (the study of topography) is
a way to decipher the dynamic of a planet. For that purpose, the statistics
of height and slopes may be described by different tools, at local
and global scale. We propose here to use the \emph{multifractal} approach
to describe fields of topography. This theory both encompass height
and slopes and other statistical moment of the field, tacking into
account the scale invariance. Contrary to the widely used \emph{fractal}
formalism, \emph{multifractal} is able to describe the intermittency
of the topography field. As we commonly observe the juxtapostion of
rough and smooth at given scale, the \emph{multifractal} framework
seems to be appropriate for hypsometric studies. Here we analyze the
data at global scale of the Earth, Mars, Mercury and the Moon and
find that the statistics are in good agreement with the \emph{multifractal}
theory for scale larger than $\sim10km$. Surprisingly, the analysis
shows that all bodies have the same \emph{fractal} behavior for scale
smaller than $\sim10km$. We hypothesized that dynamic topography
of the mantle may be the explanation at large scale, whereas the smaller
scales behavior may be related to elastic thickness.
\end{abstract}

\section{Introduction}

Scaling of coastlines was empirically studies by \citet{Richardson1961},
and \citet{Mandelbrot1967} interpreted his reuslts in terms of fractals.
Fractals are geometric sets of points that have a scale symmetry.
Geophysical examples of scaling include turbulent phenomena including
clouds, the wind, the ocean river flows, as well as various solid
earth fields including rock faults and topography. Most systems of
geophysical interest are mathematical fields, not geometric sets.
When scaling, they will generally be multifractals. A general way
to quantify this is to determine the statistical moments of fluctuations
of the field, (generalized) structure functions. Denoting the fluctuation
in the topography over a distance $\varDelta x$ by $\varDelta h$($\varDelta x$),
the $qth$ order structure function is $\left\langle \varDelta h(\varDelta x)^{q}\right\rangle $.
If the sytem is scaling, then this is a power law of the lag $\varDelta x$
: $\varDelta x^{\zeta(q)}$. The field is mono\emph{fractal} if $\zeta(q)=qH$
where H is named in honour of Hurst; in this linear case the field
is quasi-Gaussian. In the more general multifractal case, $\zeta(q)=qH-K(q)$
where $K(q)$ is a convex function with $K(1)=0$, it determines the
multifractality, the intermittency, the ``spikeness'' of the field.
Numerous studies haves shown that in several context, topography is
scaling on a significant range of scales. 

For multifractal processes, local estimates of fractal dimensions
will be different from one location to another, they will be stochastic.
It is thus possible to interpret the topography of regions with quite
different slope distributions in a unified multifractal framework.
This suggests that even a global analysis of the topography of a planet
might be scaling and multifractal despite of its diversity and complexity.
Previous studies have established that the Earth's topography is to
a good approximation multifractal over a very wide range of scales
\citep{Lavallee1993,Gagnon2006}. In the general case, $\zeta(q)$
is a concave function; in order to characterize or model multifractals
one takes advantage of the existence of stable, attractive statistical
behaviour: universality classes \citep{Schertzer1987}. 

In a previous analysis, we performed a global analysis on the topographic
MOLA data from Mars \citep{Smith2001}. We also find a good agreement
with universal multifractals but we found two scaling ranges with
different characteristics \citep{Landais2015}. The statistical structure
was found to be different at small scales (nearly monofractal) and
large scales (multifractal) with a transition occurring at around
$10km$. This behavior has been confirmed recently with other analyses
\citep{Deliege2016}.

The goal of this article is to extend this pioneering Martian work
to all planetary bodies whose topography is well estimated: the Earth,
the Moon and Mercury. There is topography data for Venus and Titan
but unfortunately too much data is missing to have a similar analysis
on the global scale.

\section{Universal multifractal theory}

We first define the fluctuations $\varDelta h$($\varDelta x$). The
simplest definition is the altitude differences, the slopes multiplied
by $\varDelta x$, the most natural indicator of roughness. But there
are many others way to define fluctuations. Wavelets provide a general
method. Indeed, their coefficients define fluctuations (with appropriate
normalization). The simple altitude difference corresponds to the
``poor man'' wavelet and can be advantageously replaced by the Haar
wavelet that is more accurate and is useful over a wider range of
exponents ($-1<H<1$, rather than $0<H<1$ for differences, see \citet{Lovejoy2014}
and paragraph below for a precise definition of Haar fluctuations. 

\paragraph{Statistical moments}

We can compute any statistical moment $M_{q}$ of order $q$ define
by:

\begin{equation}
M_{q}(\Delta x)=<\Delta h(\Delta x)^{q}>
\end{equation}

With $<>$, denoting the statistical average. If $q=2,$ it simply
correspond to the variance. In principle, every orders (even non-integer
orders) must be computed to fully revealed the whole variability of
the data. If the field is scaling, all the statistical moment are
expected to follow a power-law with scale. 

\paragraph{Multifractality}

Scaling allows us to introduce two distinct statistical processes
: monofractal and multifractal. For a detailed description of the
formalism we apply in this study, the readers can refer to \citet{Lovejoy2013a}
briefly summed up in \citet{Landais2015} . We now quickly recall
the main notions here after.
\begin{itemize}
\item In the usual gaussian monofractal case the parameters $H$ is sufficient
to describe the statistic of all the moments of order $q$ (equation
\ref{eq:monfractal}). Ther is no intermittency, meaning that the
roughness of the field is spatially homogenous despite of its fractal
variability regarding to scales. For example, the value $H=0.5$ correspond
to the classic Brownian motion. This kind of statistical object has
proved to be relevant in many local and regional analysis of natural
surfaces \citep{Orosei2003,Rosenburg2011}, at least on restricted
ranges of scales but fails to give full account to the intermittency
commonly observed on larger topographic datasets. 
\begin{equation}
M_{q}\sim\Delta x^{qH}\label{eq:monfractal}
\end{equation}
\item In the multifractal case, $H$ is no more sufficient to fully describe
the statistics of the moments of order $q$. An additional convex
function $K(q)$ depending on $q$ is required (see eq. \ref{eq:multifra}).
The moment scaling function $K$ slightly modifies the scaling law
of each moment. The consequence on the corresponding field appears
clearly on simulations \citep{Gagnon2006} : the field exhibit a juxtaposition
of rough and small places that are clearly more realistic in the case
of natural surfaces. Moreover, it is possible to restrain the generality
of the function $K(q)$, considering only universal multifractals,
a stable and attractive class proposed by \citet{Schertzer1987} for
which the multifractality is completely determined by the mean intermittency
$C_{1}=\left(\frac{dK(q)}{dq}\right)_{q=1}$ (codimension of the mean)
and the curvature $\alpha$ of the function $K$, $\alpha=\frac{1}{C_{1}}\frac{d^{2}K(q)}{dq^{2}}$
(the degree of multifractality). In that case the expression of $K$
is simply given by equation \ref{eq:universak}.
\begin{equation}
M_{q}\sim\Delta x^{qH-K(q)}\label{eq:multifra}
\end{equation}
\end{itemize}
\begin{equation}
\xi(q)=qH-K(q)\label{eq:structure}
\end{equation}

\begin{equation}
K(q)=\frac{C_{1}}{\alpha-1}(q^{\alpha}-q)\label{eq:universak}
\end{equation}

We see that the monofractal case correspond to $(H\neq0,C_{1}=0)$
or $(H\neq0,C_{1}\neq0,\alpha\rightarrow0)$. 

\section{Dataset}

The topography of a planet is defined as the difference between the
distance of the planetary surface and the geoid. For Mars \citep{Smith2001},
Mercury \citep{Cavanaugh2007} and the Moon \citep{Smith2010}, we
are used topographic data stored in PDS (http://pds-geosciences.wustl.edu)
whereas the Earth data \citep{Amante2009} are gathered from numerous
global and regional data sets. Table \ref{tab:Datasets-characteristics}
sums up the main characteristics of the datasets. Each has already
been previously analysed. 

The Earth has been studied for multifratal purpose by \citet{Gagnon2006}
using ETOPO5 dataset. They proposes to analyse separately continents
and ocean and found that $H$ is varying from 0.46 for bathymetry
and 0.66 for continent. The dataset considered in our study (ETOPO1,
\citealp{Amante2009}) is an arc-minute global relief model of the
Earth.

On Mars, the main source of topographic data is the Laser altimeter
MOLA \citep{Smith2001} that allowed to perform extensive statistical
analysis with different roughness indicators on sliding windows revealing
interesting correlation with geological units \citep{Aharonson2001,Kreslavsky2000}.
The monofractal scaling of the topography of Mars has also been studied
by \citet{Orosei2003} through the local computation of the scale
independent Hurst parameters revealing a high disparity of values
across the martian surfaces as expected for multifractal topography.

On the Moon, the high-precision topographic data obtained by the laser
altimeter LOLA \citep{Smith2010} has been extensively used. \citet{Kreslavsky2013}
computed maps of roughness at hectometer and kilometer scales revealing
poor correlations between these two scales. Moreover, \citet{Rosenburg2011}
measured $H$. They not only identified a transition that occurs around
1km at most location but they also found significantly different values
of $H$ in the Highlands ($H=0.95$) and in the Marias ($H=0.76$). 

On Mercury, by using the MLA data \citep{Cavanaugh2007}, \citet{Pommerol2012}
computed roughness indicators on extracted profiles from geologically
distinct regions. Due to the eccentricity of the orbit, only the northern
hemisphere could be mapped by laser altimetry with a resolution of
about 5 kilometers. The use of pairs of stereoscopic images has finally
made possible to develop an overall map of the topography of Mercury
\citep{solomon2001messenger,hawkins2007mercury}. We analyzed both
MLA data only and the full map (from both laser and stereoscopy) and
found no significant difference, except the fact that larger scale
are available for the full map. We thus choose to present here the
full map only. This result confirm that there is no significant bias
(at least with on multifractal properties) between stereoscopic and
laser altimetric techniques.

The current study proposes to extend the scope of the multifractal
analysis already performed on Earth and Mars \citep{Gagnon2006,Landais2015}
to all the bodies in the solar system for data is adequate . Thus
the case of Earth, Mars, Moon and Mercury will be comparatively discussed.
The case of Venus is not considered here despite of the existence
of a dataset collected by Magellan because of the relative lack of
topographic data comparing to the other bodies \citep{Ford2014}.
The case is equivalent for Titan \citep{Stiles2009}.

\begin{table}[b]
\begin{centering}
{\small{}}%
\begin{tabular}{|c|c|c|c|c|c|c|c|c|}
\hline 
 & {\small{}source} & {\small{}radius (km)} & {\small{}Resolution} & {\small{}min scale} & {\small{}max scale} & {\small{}lines} & {\small{}columns} & {\small{}nb fluctuations }\tabularnewline
\hline 
\hline 
{\small{}Earth} & {\small{}\href{https://www.ngdc.noaa.gov/mgg/global/global.html}{ETOPO1}} & {\small{}6371} & {\small{}60 pix/deg} & {\small{}1 853 m} & {\small{}20 015 km} & {\small{}10 800 } & {\small{}21 600} & {\small{}0.2 billions}\tabularnewline
\hline 
{\small{}Mars} & {\small{}\href{https://www.ngdc.noaa.gov/mgg/global/global.html}{MOLA}} & {\small{}3390} & {\small{}128 pix/deg} & {\small{}462 m} & {\small{}10 650 km} & {\small{}22 528} & {\small{}46 080} & {\small{}1 billion}\tabularnewline
\hline 
{\small{}Moon} & {\small{}\href{http://pds-geosciences.wustl.edu/missions/lro/lola.htm}{LOLA}} & {\small{}1737} & {\small{}512 pix/deg} & {\small{}60m} & {\small{}5 457 km} & {\small{}92 160} & {\small{}184 320} & {\small{}13 billions}\tabularnewline
\hline 
{\small{}Mercury} & {\small{}\href{http://pds-geosciences.wustl.edu/missions/messenger/mla.htm}{MLA}} & {\small{}2240} & {\small{}64 pix/deg} & {\small{}665 m} & {\small{}7 037 km} & {\small{}11 520} & {\small{}23 040} & {\small{}0.0002 billion}\tabularnewline
\hline 
\end{tabular}
\par\end{centering}{\small \par}
\centering{}\caption{Characteristics of the datasets\label{tab:Datasets-characteristics}}
\end{table}

\section{Methodology}

In our previous analysis \citep{Landais2015}, we considered the 1-D
topographic profiles directly extracted from the along-track measurement
of MOLA stored in PDS (http://pds-geosciences.wustl.edu, \citealt{Smith2001}).
As the data are irregularly sampled due to the presence of clouds
and instrument problems, we used multifractal simulations to study
the effect of a MOLA-like irregular signal on the Haar fluctuations
. It turned out that, most probably due to the small fraction of missing
data, the irregularity had no detectable impact on the analysis. We
also found that the use of the gridded data also produced the same
results has the direct use of the more reliable along-track measurement,
the conclusion being that for the purpose of a global analysis, the
extrapolated gridded map for each body is sufficient to recover the
global statistical parameters. The methodology used here is therefore
much simpler and only relies on the gridded data . We only considered
1D North-South profiles and computed the Haar fluctuations at different
lags $\Delta x$. The simplification to 1-D is reasonable as we perform
a global statistical analysis. In addition, the North/South direction
is more relevant than East-West because each profile as the same length.
Figure \ref{fig:4-profiles,-1} provides an example of 1-D profiles
extracted from the gridded field for each body. See \citet{Landais2015}
for a review of the different biases that could result from such an
approach.

We implicitly consider that the global statistic are isotropic. This
assumption is reasonable for the purpose of a global analysis given
the fact that shape of various orientation can be found on a given
body. Although local anisotropy is commonly observed \citep{kreslavsky2003north,bondarenko2006north,bills2014harmonic},
we assume it is erased by the spatial averaging. Isotropic multifractal
processes readily produce strong local anistropy so that the question
of systematic scale dependant statistical anisotropy is not easy to
establish. Anisotropy remains an important issue and will be more
carefully considered in future works

\begin{figure}
\subfloat[\label{fig:4-profiles,-1}]{\includegraphics[width=0.45\textwidth]{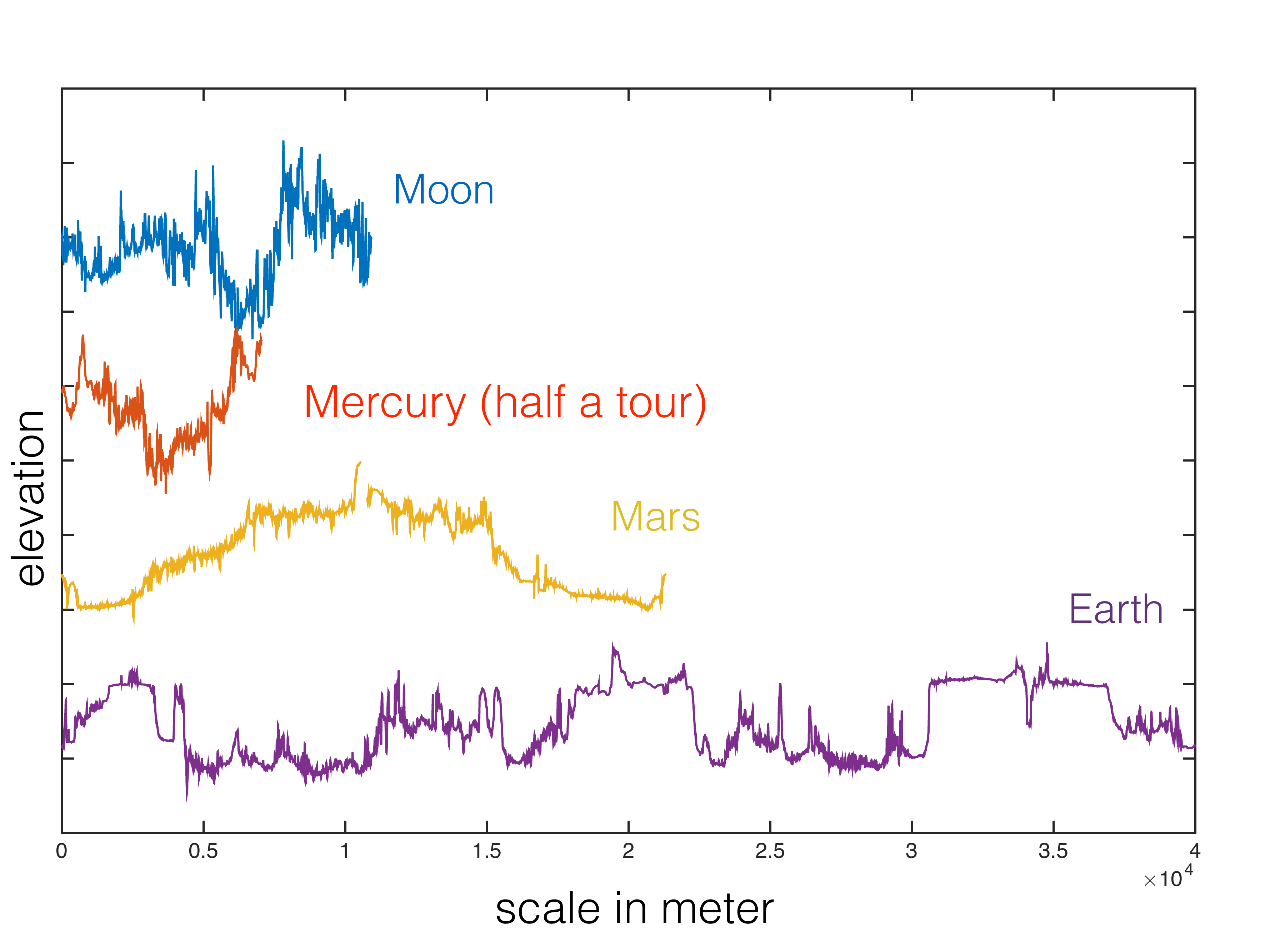}

}\hfill{}\subfloat[\label{fig:Definition-of-the}]{\includegraphics[width=0.45\textwidth]{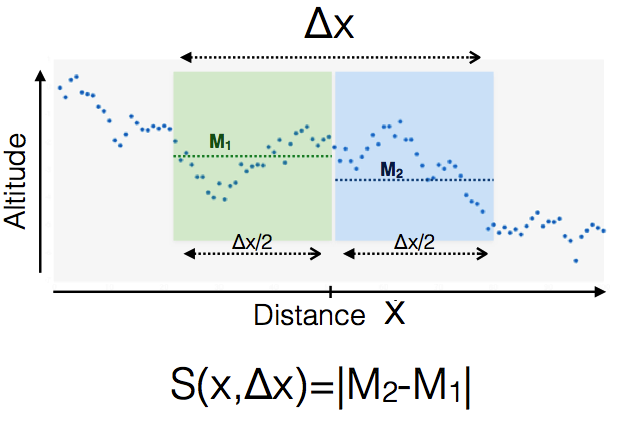}

}\caption{(a) Topographic dataset: 4 typical profiles, 1 for each body in the
scope of this study. The length corresponds to 1 complete circumference
of the planets except in the case of Mercury (only half a circumference)
(b) Definition of the Haar fluctuation used to perform the statistical
analysis. $M_{1}$ (respectively $M_{2}$) is the average of the first
half (respectively second half) of the topographic profile.}
\end{figure}

\paragraph{Haar fluctuation and statistical moments}

At a given location $x$ and a given scale $\varDelta x$ corresponding
to $N$ successive elevation data on the grid, we average separately
the first $\frac{N}{2}$ points $M_{1}(x,\varDelta x)$ and last $\frac{N}{2}$
points $M_{2}(x,\varDelta x)$. The Haar fluctuation is simply defined
as the difference $S(x,\varDelta x)=\left|M_{2}(x,\varDelta x)-M_{1}(x,\varDelta x)\right|$.
This definition is illustrated by figure \ref{fig:Definition-of-the}.
The Mean Haar Fluctuation ($MHF$, moment of order 1) is simply obtained
by averaging all the available haar fluctuations in a dataset. By
extension, other statistical moments of any order q, $M_{q,HF}$ may
be computed by averaging the fluctuations raised to the power of q
:

\[
MHF=<S(x,\varDelta x)>
\]

\[
MHF_{q}=<S(x,\varDelta x)^{q}>
\]

\section{Results for Earth, Mars, Mercury and the Moon}

\paragraph*{Mean Haar Fluctuations ($MHF$)}

\begin{figure}[h]
\includegraphics[width=1\textwidth]{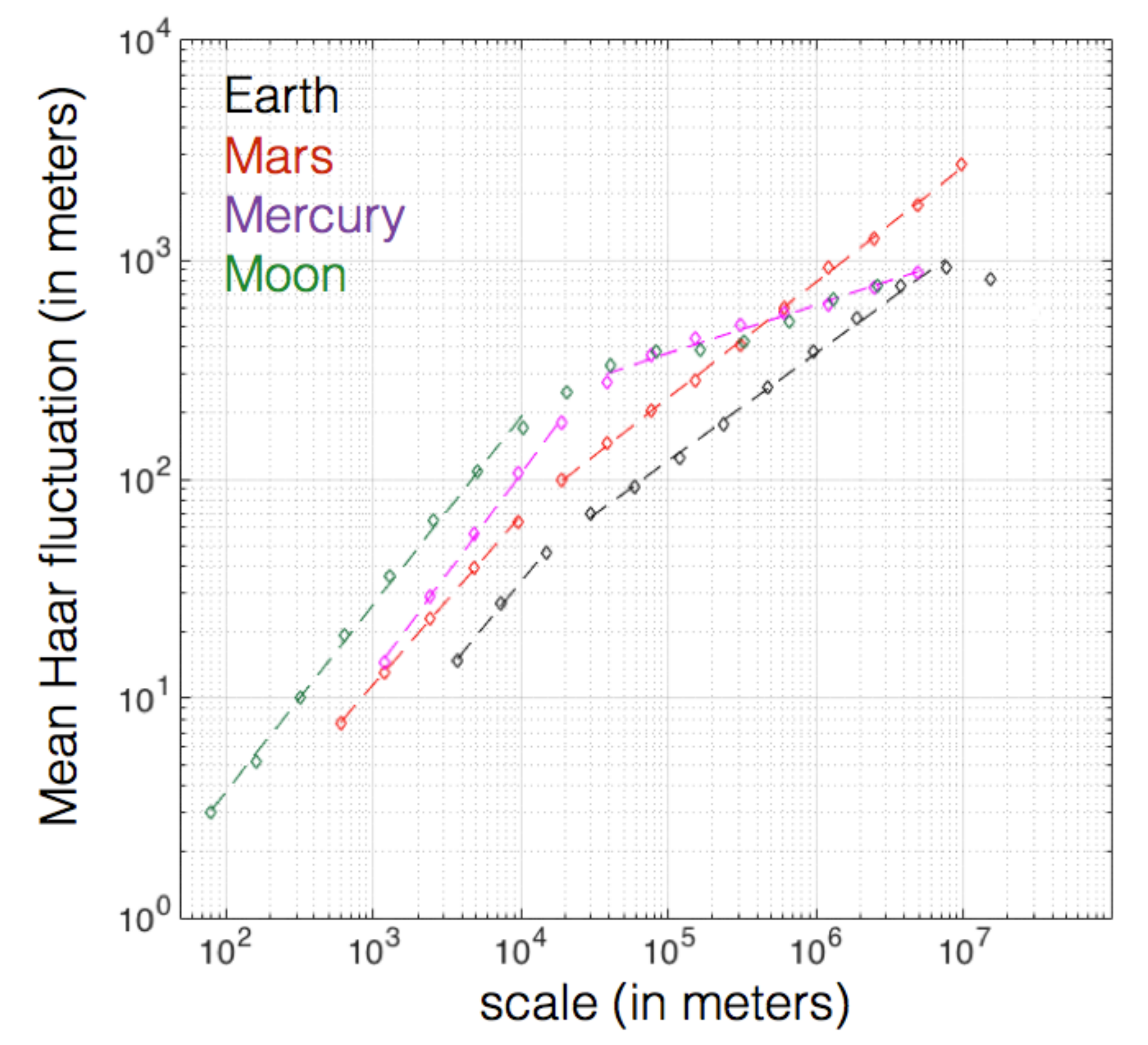}

\caption{Mean Haar fluctuations $MHQ$ (order 1) as a function of scales for
the 4 planetary bodies. \label{fig:Mean-Haar-fluctuations}}
\end{figure}

Figure \ref{fig:Mean-Haar-fluctuations} shows the Mean Haar Fluctuations
($MHF$, moment of order 1) for each body on a log-log plot. One can
observe its scaling behavior as a function of the real distance (in
meters). One can observe that at small scale, $MHF$ is from the larger
to the smaller for : the Moon, Mercury, Mars, the Earth. This simply
means that statistically, the roughness is from the larger to the
smaller: for the Moon, Mercury, Mars, the Earth. Thus a astronaut
(coming from the Earth) would experience topography differently by
looking at the landscape of other planetary bodies. He would feel
smaller in front of a rougher landscape at his/her scale. This feeling
should be larger for the Moon. Another interesting features is the
resemblance between i) the curves of Earth and Mars and ii) the curves
the Moon and Mercury. For the two small bodies, the $MHF$ is clearly
above the two others, except at the highest scales. By interpreting
it as a roughness indicator, this feature simply reflects the well-known
high level of roughness of small bodies, a consequence of intense
craterisation shaping their surfaces. 

As expected, the global $MHF$ increases with scale in all cases,
simply reflecting the fact that larger scales yield larger differences
in elevation. Nevertheless for the Earth and Mercury, at large scales,
the $MHF$ begins to decrease before reaching its maximum scale. More
specifically, as our goal is to study the global scaling behavior
of topography, we expect this global increase of the $MHF$ to be
linear on a log-log plot. It is clearly not the case on the entire
available range of scales. Still noticeable scaling appears but over
restricted ranges of scale : a transition seems to occur, separating
two distinct scaling regimes. Such a transition is observed for all
the bodies and interestingly, it occurs around 10-20 km in each case
including the Moon. The nature of this transition discussed in our
previous analysis focused on Mars and already pointed out by other
authors in the case of Mars \citep{Malamud2001}, remains unknown. 

\begin{figure}
\includegraphics[width=1\textwidth]{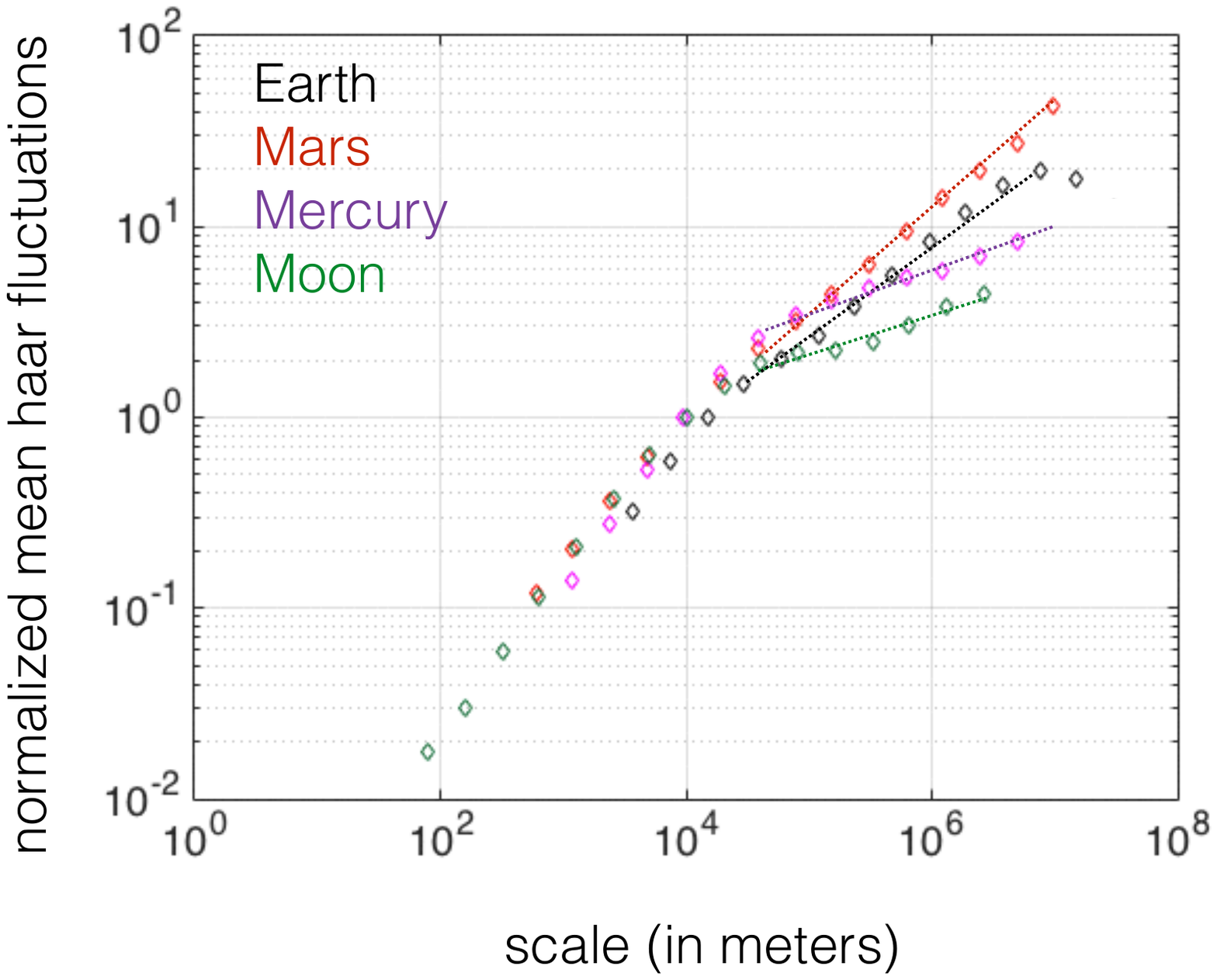}\caption{Mean Haar fluctuations normalized in order to be approximatively equal
at scale 10km, as a function of scale. The normalization does not
modify the scaling behavior from Fig. \ref{fig:Mean-Haar-fluctuations}
but emphasize the transition that seems to occur around 10km. \label{fig:Mean-normalized} }
\end{figure}

On Figure \ref{fig:Mean-normalized}, the $MHF$ are normalized by
their respective values around 10 km in order to emphasize the transition
at that scale. As one can see, the slope at small scales ($<10km$)
are rather similar ($H\sim0.8)$ whereas significantly different slopes
are observed at large scales ($H\sim0.2-0.5$). The scaling is excellent
at large scales in the case of Mars and good in the case of Earth
and Mercury. In the case of the Moon, data points are more dispersed,
and $H$ might be less well defined. Note also at small scales, the
available range for Earth and Mercury is limited and might result
in an unconvincing fit. The values obtained for $H$ for each body
by computing a linear regression on the distinct ranges of scale is
reported on table \ref{tab:Estimations-of-fractal-multifractal-parameters}. 

\paragraph*{Statistical moments $MHF_{q}$}

\begin{figure}
\includegraphics[width=1\textwidth]{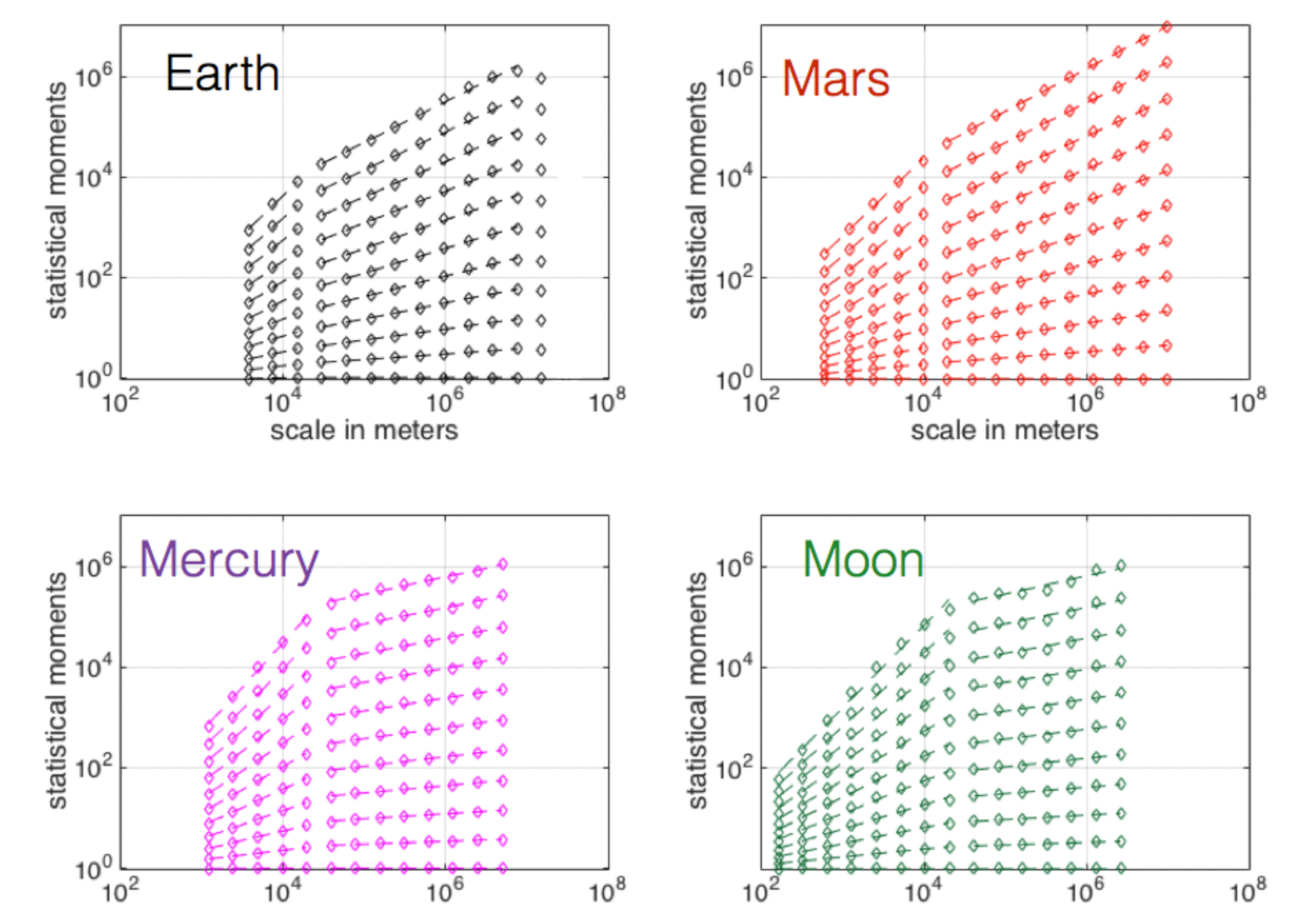}

\caption{Plot of different statistical moments for the four bodies. The average
fluctuations from figure \ref{fig:Mean-Haar-fluctuations} are shown
as diamond whereas the moments of order 2 (average squared fluctuations)
are triangles (top curves). In between are the non-integer order statistical
moments. Although 21 moments have been computed for the purpose of
this analysis (from 0.1 to 2 by steps of 0.1), only a few non-integers
moments are plotted here for order 0.1 to order 2 by step of 0.2 \label{fig:Plot-of-different-1}}
\end{figure}

\begin{figure}
\includegraphics[width=1\textwidth]{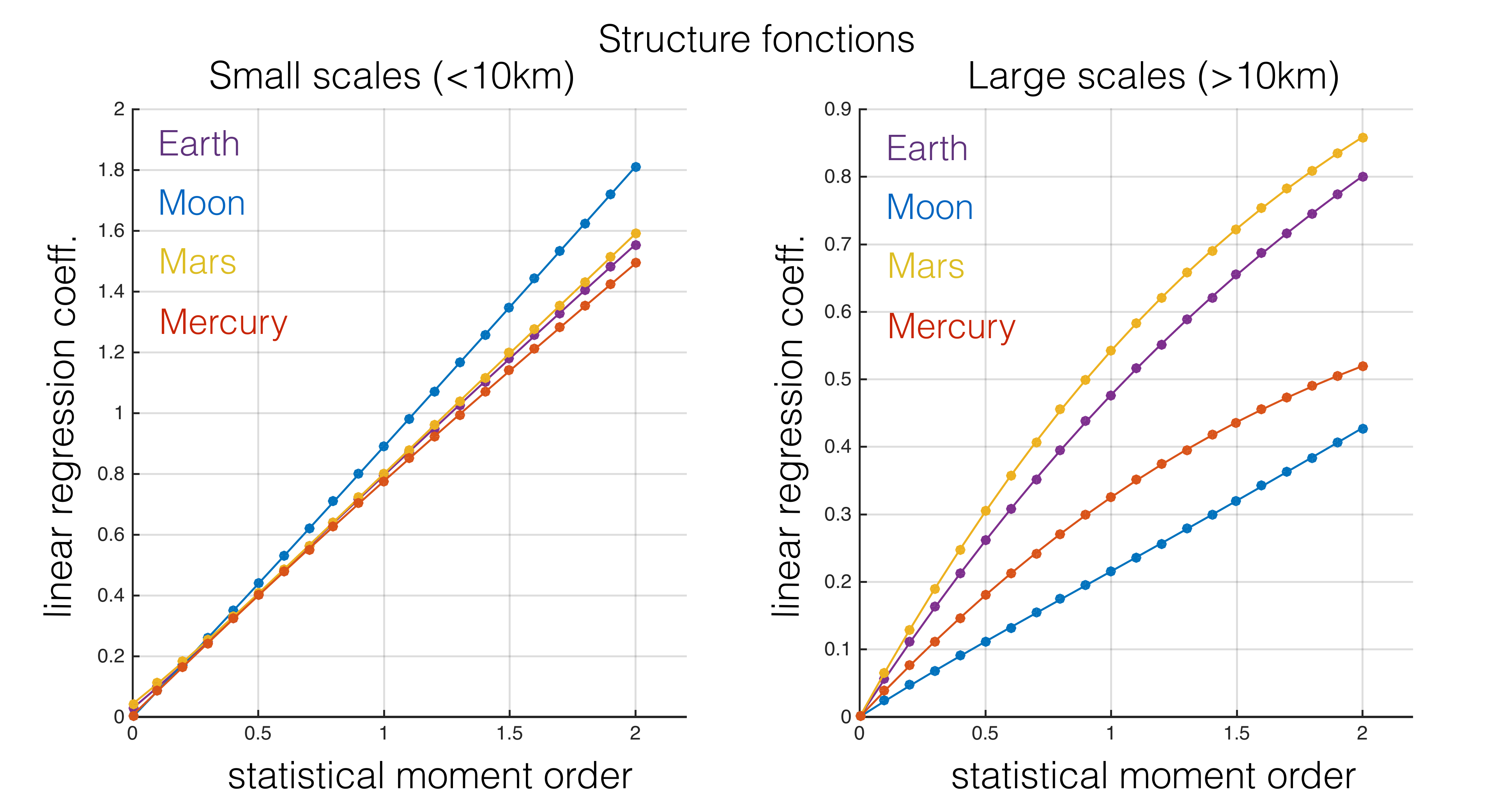}

\caption{Structure function for different ranges of scales\label{fig:Structure-function-for} }
\end{figure}

In the case of universal multifractals, all the statistical moment
will scale according to equation \ref{eq:multifra}, the $MHF$ being
the particular case for which $q=1$. Thus we can estimate the other
multifractal exponents by computing statistical moments $MHF_{q}$
. On Figure \ref{fig:Plot-of-different-1}, the $MHF_{q}$ are plotted
for different values of q and for the different bodies. The next step
is to compute linear regressions on every curve and on the distinct
identified scaling regime. The log-log slopes $\zeta(q)$ may then
be plotted as a function of $q$ for each body and for each range
of scale (see Figure \ref{fig:Structure-function-for}) in order to
visualize the function $\xi$ defined by equation \ref{eq:structure}.
A linear $\xi(q)$ is the signature of monofractality whereas a curved
$\xi(q)$ indicates a multifractal behavior according to equation
\ref{eq:monfractal} and \ref{eq:multifra}. Interestingly, Figure
\ref{fig:Structure-function-for} clearly shows that on the distinct
scaling regimes (low scale and large scales) the behavior is significantly
different. On the range of small scales ($<10km$, plot on the left),
the curves are rather similar for all the four bodies and very close
to straight lines indicating that the statistics found to be roughly
monofractal (small $C_{1}$) over the range. Over the range of large
scales ($>10km$, plot on the right), we obtained curved structure
functions in most cases revealing the multifractal nature of the statistics
of topography over the range. The multifractal parameters are computed
according to equation \ref{eq:multifra} and reported on table \ref{tab:Estimations-of-fractal-multifractal-parameters}.
Whereas the case of Mars, Mercury and Earth are similar of value of
$C_{1}$ around 0.1, the case of the moon seems to be an exception
with weak multifractal properties over that range of scales ($C_{1}$
close to 0).

\begin{table}[b]
\begin{centering}
\begin{tabular}{|c|c|c|c|c|}
\hline 
scale<10km & \multirow{1}{*}{Earth} & Mars & Moon & Mercury\tabularnewline
\hline 
\hline 
H & \multirow{1}{*}{0.823 \textpm{} 0.004} & 0.773 \textpm{} 0.003 & 0.878 \textpm{} 0.002 & 0.922 \textpm{} 0.003\tabularnewline
\hline 
$\alpha$ & NA & NA & NA & NA\tabularnewline
\hline 
$C_{1}$ & 0.01 \textpm{} 0.01 & 0.02 \textpm{} 0.006 & 0.02 \textpm{} 0.04 & 0.026 \textpm{} 0.005\tabularnewline
\hline 
\end{tabular}
\par\end{centering}
\begin{centering}
\begin{tabular}{|c|c|c|c|c|}
\hline 
scale>10km & \multirow{1}{*}{Earth} & Mars & Moon & Mercury\tabularnewline
\hline 
\hline 
H & \multirow{1}{*}{0.479 \textpm{} 0.001} & 0.53 \textpm{} 0.001 & 0.226 \textpm{} 0.002 & 0.248 \textpm{} 0.002\tabularnewline
\hline 
$\alpha$ & 1.70 \textpm{} 0.08 & 1.80 \textpm{} 0.06 & 1.4 \textpm{} 0.1 & 1.85 \textpm{} 0.1\tabularnewline
\hline 
$C_{1}$ & 0.093 \textpm{} 0.002 & 0.110 \textpm{} 0.002 & 0.03 \textpm{} 0.01 & 0.059 \textpm{} 0.002\tabularnewline
\hline 
\end{tabular}
\par\end{centering}
\centering{}\caption{Estimations of multifractal parameters\label{tab:Estimations-of-fractal-multifractal-parameters}.
NA stands for non-applicable. If the value of $C_{1}$ is very small
(here <0.02), we can consider that the field is not multifractal and
the value of $\alpha$ is not interpretable.}
\end{table}

\section{Discussion and conclusion}

By averaging the fluctuations at different scales, we have revealed
global statistical pattern of planetary bodies. \foreignlanguage{english}{We
tested and validated the multifractal approach on the four bodies
with empirically well estimated topography : the Earth, Mars, Mercury
and the Moon. We found that a transition occurs at about 10 km and
that it is a general property of all planetary topographies. Below
10 km, differences in altitudes decrease more rapidly when the scale
decreases. }

As suggested by \citet{araki2009lunar} and \citet{nimmo2011geophysical},
for the transition in the power spectral density, we propose the interpretation
that the elastic thickness of the lithosphere is responsible for this
transition by acting against the deformations caused by the different
surface processes in two regimes. At scales smaller than the elastic
thickness $T_{e}$, a modification of the surface (for example, following
an impact) does not make it possible to generate isostatic compensation.
The new relief can therefore remain present. The slopes of neighboring
facets tend to be correlated with each other and give rise to fluctuations
in topography rapidly increasing with the scale (structured aspect,
high $H$). The relief profile tends to be constructive since the
slopes are highly correlated. At scales higher than $T_{e}$, a change
in relief triggers an isostatic compensation which tends to oppose
the large variations of the relief. The slopes of neighboring facets
tend to be anti-correlated and the topographic profile is rougher.
The relief oscillates around a mean value since the slope are more
anti-correlated. In this configuration, the altitude fluctuations
increase only slightly regarding to scale (low $H$). The common transition
could be explained by the averaged value of the elastic thickness
quite similar for the 5 bodies \citep{grott2008evolution,barnett2000elastic,nimmo2004depth}.

At scales larger than 10 km, all planetary bodies are different. Interestingly,
the scaling law is characterized for the Moon by $H=0.2$, Mercury
by $H=0.3$, Mars and Earth by $H=0.5$. The smaller the body, the
less intense its internal activity due to intense thermal cooling.
The value of $H$ may be related to its geological activity. One can
speculate that a more intensively convecting mantle yields a higher
value of $H$. This explanation links the large scale with dynamic
topography \citep{Hager1985}. The fact that only large scale topography
is strongly multifractal is coherent with this explanation because
multifractal behavior is related to fluids mechanics and turbulent
scale cascade. The geological origin of this transition will be investigated
in future works. 

From our result, this pattern seems coherent on large ranges of scale
throughout the different bodies. Although suggesting that a few processes
might operate simultaneously at different scales, this result is not
incompatible with the existence of process operating at a specific
altitude or locations. For example the ``glacial buzz'' saw effect
seems limit the presence of high altitude on the Earth only \citep{lorenz2011hypsometry}.
Our results simply suggest that the contribution of such process to
global statistics can be neglected because if a strong altitude dependent
process occurs, it should have broken the scaling behavior.

As a future work, we plan to perform local analysis on area defined
by geological boundaries or altitude level to better understand the
link between the scaling behavior of topography and natural processes
operating at different location and altitude. 

\section*{Acknowledgment}

We acknowledge support from the ``Institut National des Sciences
de l'Univers'' (INSU), the \textquotedbl{}Centre National de la Recherche
Scientifique\textquotedbl{} (CNRS) and \textquotedbl{}Centre National
d'Etudes Spatiales\textquotedbl{} (CNES). This work was supported
by the Programme National de Plan\'etologie (PNP) of CNRS/INSU, co-funded
by CNES 

\bibliographystyle{agu04}

\section*{Annex : Bayesian regression}

In order to estimate the best set of parameters $(H,\,C_{1},\alpha)$
modeling the data, the parameters can be estimated in a classical
way by performing regressions on the function $\zeta$ near the mean
$(q=1$) to quantify its curvature related to $\alpha$ and $C_{1}$
\citet{Lovejoy2013a} by the theoretical formulas:

\begin{equation}
H=\left.\frac{d\xi(q)}{dq}\right|_{q=1}\quad C_{1}=\left.\frac{dK(q)}{dq}\right|_{q=1}\quad\alpha=\left.-2\frac{d^{2}K(q)}{dq^{2}}\right|_{q=1}/\left.\frac{dK(q)}{dq}\right|_{q=1}\label{eq:parameter_fit}
\end{equation}

As a reminder, the scaling exponents $\zeta(q)$ are themselves the
products of linear regression, so the fits from eq. \ref{eq:parameter_fit}
are only indirectly related the data. We wish to avoid this method
which will not make possible to judge the quality of the estimates,
especially as the multifractal component is rather weak when the function
$\zeta(q)$ is only weakly curved. 

We propose a new approach based on principle of Bayesian inversion
\citep{tarantola1982generalized} which allows to construct a posterior
probability distribution of the parameters (mean, most probable value,
standard deviation) from observations. In practice, these distributions
can be estimated iteratively by applying the Metropolis rule to construct
a Monte Carlo Markov chain \citep{mosegaard1995monte} containing
the different sets of parameters. We summarize the main lines of this
technique, already used on photometry problems \citep{schmidt2015realistic,Fernando_SurfacereflectanceMars_JoGRP2013}.
As a first step, it is necessary to evaluate the quality of the individual
linear regressions of each moment. In this step, we attribute to each
point an empirical uncertainty with a centered gaussian distribution.
The latter will be the higher as the linear correlation through the
scale is accurate. Then, we tested the direct model , computed by
applying the laws of equations \ref{eq:structure} and \ref{eq:structure},
for different random set of parameters $(H,\,C_{1},\alpha)$. Synthetic
realizations are then compared to observations. The Monte Carlo Markov
chain is created according the metropolis rules, using the likelyhood
of empirical uncertainties. . This method allow us to estimated realistic
uncertainty bars on parameter $(H,\,C_{1},\alpha)$, from the observational
data (see table \ref{tab:Estimations-of-fractal-multifractal-parameters})%

\end{document}